\documentclass[conference]{IEEEtran}
\usepackage{cite}
\usepackage{amsmath,amssymb,amsfonts}
\usepackage{algorithmic}
\usepackage{graphicx}
\usepackage{textcomp}
\usepackage{xcolor}
\usepackage[hyphens]{url}
\usepackage{amssymb}
\usepackage{pifont}

\def\BibTeX{{\rm B\kern-.05em{\sc i\kern-.025em b}\kern-.08em
    T\kern-.1667em\lower.7ex\hbox{E}\kern-.125emX}}

\title{ACiS: Complex Processing in the Switch Fabric}
\author{Pouya Haghi$^\ast$~~~Anqi Guo$^+$~~~Tong Geng$^\ast$~~~Anthony Skjellum$^\#$~~~Martin Herbordt$^+$ \\
$^\ast$University of Rochester~~~~$^+$Boston University~~~~$^\#$Tennessee Tech
}

\begin{document}
\maketitle
\thispagestyle{plain}
\pagestyle{plain}

\begin{abstract}
For the last three decades a core use of FPGAs has been for processing communication: FPGA-based SmartNICs are in widespread use from the datacenter to IoT. Augmenting \textit{switches} with FPGAs, however, has been less studied, but has numerous advantages built around the processing being moved from the edge of the network to the center. Communication switches have previously been augmented to process collectives, e.g., IBM BlueGene and Mellanox SHArP, but the support has been limited to a small set of predefined scalar operations and datatypes.

In this manuscript we present an outline of our work so far on ACiS, a framework and taxonomy for \textit{Advanced Computing in the Switch} that unifies and expands our previous work in this area. In addition to fixed scalar collectives in current use (Type 1), we propose three more types of in-switch application processing: (Type 2) \textit{User-defined} operations and types, including data structures; (Type 3) \textit{Look-aside} operations that have state within the operation and can have loops; and (Type 4) \textit{Fused} collectives built by fusing multiple existing collectives or collectives with map computations. ACiS is supported in hardware with modular switch extensions including a CGRA architecture. Software support for ACiS includes evaluation and translation of relevant parts of user programs, compilation of user specifications into control flow graphs, and mapping the graphs into switch hardware. The overall goal is the transparent acceleration of HPC applications encapsulated within an MPI implementation.\footnote{This work was presented at the 2nd Workshop on FPGA Technologies for Adaptive Computing (FTAC 2024) held in conjunction with the 38th International Conference on Supercomputing (ICS 2024), Kyoto Japan, June 4, 2024. Submissions were reviewed but access restricted to workshop attendees.}
\end{abstract}

\section{Introduction}
\label{sec:intro}

High performance computing (HPC) systems are facing a crisis in performance-portability and scalability. Problems include increasing communication latency, networks that are both overloaded and underutilized (depending on the application mix), load balancing, and process skew. In this work we propose ACiS, a framework and taxonomy for \textit{Advanced Computing in the Switch}. ACiS adds application-level computation to communication switches; we find that extremely high-value computing can be enabled with minimal redesign of either network, NIC, or processing node, and that ACiS hardware can be added to the switch without changing standard dataplane architecture or loss of non-ACiS performance.

Thirty years of codesign have made FPGAs ideal communication processors. There are at least three aspects to this: (i) compute capability that is high performance and per-application configurable; (ii) communication capability that is vast and flexible through scores of multigigabit transceivers (MGTs); and (iii) tight coupling of computation with communication that enables application-level transfers in just a few cycles. Use of FPGAs in ACiS is therefore a plausible approach for addressing HPC scalability problems.

Offload of collective processing into SmartNICs \cite{in_nic_collective2, Bayatpour2021} is well established and has a number of benefits: first, it enables the bypassing of layers in the communication software stack; second, the hardware implementations are substantially faster than the software; third, it frees up the host processor for other tasks and, potentially, enables better communication-computation overlap; and fourth, some network-host communication is removed as the NIC handles additional send/receive operations. While SmartNICs are valuable, this scheme still forces processing into the endpoints. 

Another approach is to offload collective processing into the switches \cite{Graham2016, DeSensi21}. This has two additional benefits: first, latency is improved as computation is distributed rather than performed in a single source (broadcast) or endpoint (reduction); and, second, communication volume may be drastically reduced as messages are quickly merged (reduction) or slowly replicated (broadcast).  Communication switches have previously been augmented to process collectives, e.g., the IBM BlueGene project and the Mellanox SHArP switch, but the support has been limited to a small set of predefined scalar operations and datatypes (e.g., \cite{Faraj09,Graham2016}). Moreover, beyond collectives there are additional acceleration opportunities. 

With ACiS we unify and expand our previous work \cite{Haghi22,Haghi23,Haghi23b,Haghi24} to provide a framework that further augments switches to accelerate additional and more complex functions that integrate communication with computation. In addition to simple collectives (Type 1), we propose three more types of in-switch application processing: Type 2 -- \textit{User-defined} operations and types, including simple data structures; Type 3 -- \textit{Look-aside} operations that have state within the operation and can have loops; and Type 4 -- \textit{Fused} collectives built by fusing multiple existing collectives or collectives with computations. 

Implementations with reconfigurable logic have several inherent advantages over fixed logic implementations.\footnote{We note that software implementations are out of the question since they could never keep up with current line rates.} First, they are not limited to a small, fixed set of operations; second, hardware resources can be configured to match application requirements; third, support can be extended beyond simple datatypes to higher order structures such as matrices, tensors, and user defined datatypes; fourth, since reconfiguration time is similar to program load time, only resources that will actually be used need to be configured. 

\begin{figure*}[htbp]
\centering
\includegraphics[width=0.9\linewidth]{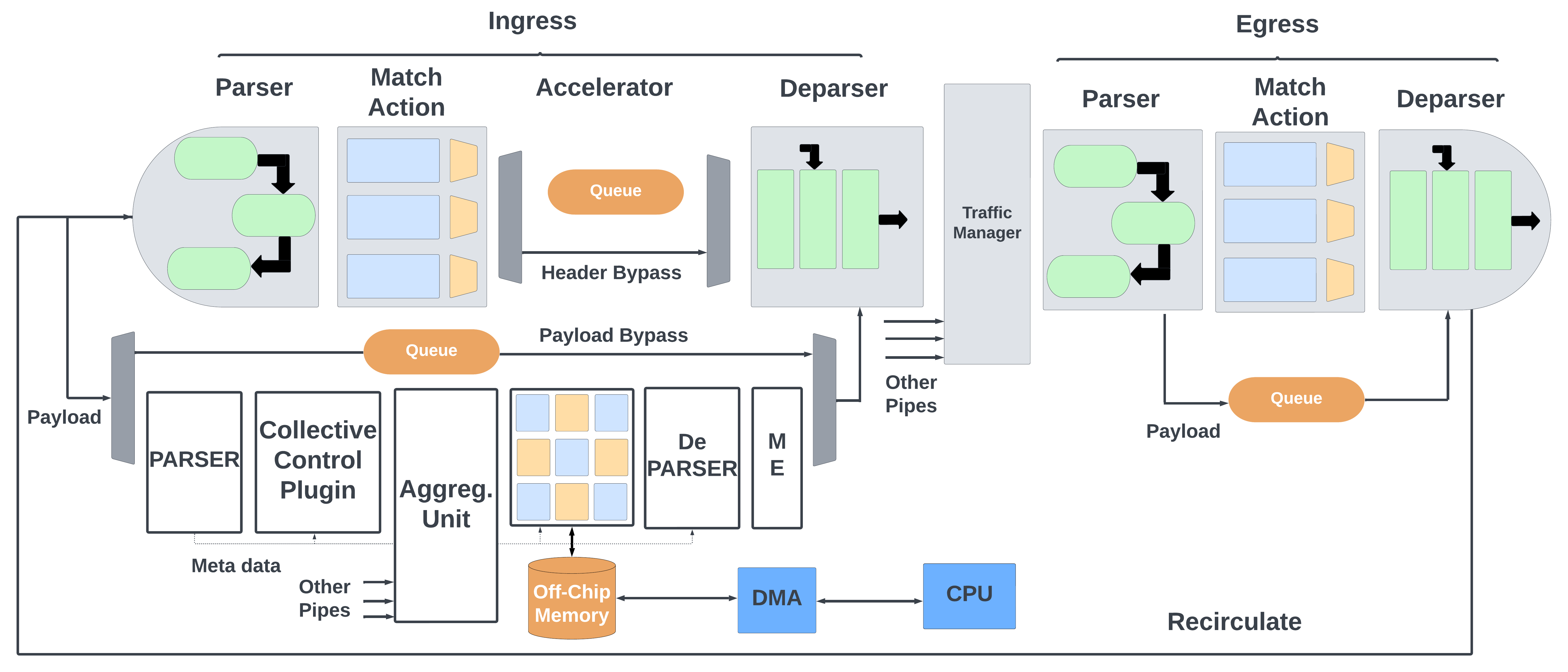}
\vspace{-0.1truein}
\caption{\small The well-known Protocol Independent Switch Architecture (PISA) enhanced with ACiS accelerator, a composable plugin to a packet processing pipeline (one pipe shown). White blocks are proposed and grey blocks are in existing switches.}
\label{fig:model-Type1to4}
\vspace{-0.05truein}
\end{figure*}

Our hardware approach begins with a seamless integration into a general router, added hardware support for aggregating (reordering) packets for reduce-type (gather-type) operations, and the flexibility to add more computational resources as the switch bandwidth increases. Complexity is supported with a CGRA-like architecture. Software support for ACiS includes evaluation and translation of relevant parts of user programs, compilation of user specifications into control flow graphs, and mapping the graph into switch hardware. Another core part of ACis is encapsulation within an MPI implementation for transparent acceleration of HPC applications. Efficacy of ACiS has been demonstrated \cite{Haghi22,Haghi23,Haghi23b,Haghi24} on a number of HPC applications including the NAS parallel benchmarks, two MiniApps from Mantevo (miniFE and HPCCG), graph convolutional networks (GCNs), PGEMM, and AMG. 

For the rest of this report we first expand on the ACiS Types, then give an overview of the hardware aspects of ACiS, followed by some experimental results, an overview of software support, and related work.

\section{ACiS Types}
\label{sec:ACiS}

We classify ACiS capabilities into progressively complex types of in-switch computing (Table~\ref{tab:reference}):

\begin{table}[!t]
  \centering
\caption{ACiS Types Reference}
\vspace*{-0.05truein}
\label{tab:config}
\begin{tabular}{|c|c|}
\hline
Type 0  & Processing single data streams      \\ \hline
Type 1  & Collectives on primitive types  \\ \hline
Type 2  & User defined operations and types  \\ \hline
Type 3  & Look-aside processing - loops and memory    \\ \hline
Type 4  & Fused collectives and map functions  \\ \hline
\end{tabular}
  \label{tab:reference}
\vspace{-0.in}
\end{table}

\noindent
{\bf Type 0:} Consists of transformations on {\bf streams} such as changes in data types or appending a CRC code. This is a well-understood subset of work proposed here and only mentioned for completeness.

\vspace*{0.04truein}
\noindent
{\bf Type 1:} Adds to Type 0 by supporting {\bf collectives}, but on a limited number of primitive data types (e.g., int) and operations (e.g., add, max). Datatypes, operations, and communication contexts (MPI communicator) are fixed. 

\vspace*{0.04truein}
\noindent
{\bf Type 2 \cite{Haghi22}:} Adds to Type 1 by supporting collectives of {\bf user-defined} operations (e.g., dot product), datatypes (e.g., matrices, sparse datatypes), and communication context with MPI communicators scalable to a large number of nodes. 

\vspace*{0.04truein}
\noindent
{\bf Type 3 \cite{Haghi23}:} Adds to Type 2 by supporting {\bf look-aside} capability: functions requiring loops and off-chip memory. The data plane has direct access to off-chip memory for storing/retrieving packet data. Examples include accelerating compression algorithms and communication-intensive parts of machine learning inference for large-scale datasets. 

\vspace*{0.04truein}
\noindent
{\bf Type 4: \cite{Haghi24}} Adds to Type 3 by supporting {\bf fused collectives} and the combining of collectives with map operations. An example is fusing AllReduce with AlltoAll found in the NAS sort benchmark. Another example is accelerating MapReduce type of operations (used in DNN inference).

\section{Related Work, ACiS Differentiation}
\label{sec:related}

Although \textit{the network is the computer} is an old trope \cite{Perry19}, in practice it has most often meant use of the network as a passive, transparent conduit. Recent years have seen the flipping of the data center model from compute-centric to data-centric \cite{Yoshida20,Morgan21}, in part, through the emergence of SmartNICs \cite{Arap16, Graham16, Bayatpour21, Xilinx21Alveo, Xilinx21SN1000, Inventec21C5020X, Silicom21N5010, Napatech21FPGASmartNICs, Intel21D5005, Mellanox21Innova} and similar network-facing devices \cite{Caulfield16}, and their use in offloaded application \cite{Chung19,Xiong19,Guo22,Guo22a,Guo23} and system \cite{Li16,Firestone18} processing.
Simple application-stream processing for collectives has also been implemented in switches by IBM and Mellanox \cite{Almasi05,Graham16,Graham20}, but these operations have had limited scope, e.g., reductions only, and on a small set of operations and simple data types (part of ACiS Type 1). Also, in general, they have offered only modest benefit in typical environments \cite{Hoefler09,Haghi21}. Our recent work has demonstrated the viability and benefit of extending in-switch processing in two ways: expanding in-switch processing support along several dimensions (Types 2-4 \cite{Haghi22,Haghi23,Haghi24}) and network-centric acceleration of specific applications \cite{Haghi20,Haghi20b}. The work described here proposes extensions and generalizations to these approaches, namely, to advance intelligent communication by augmenting switches to enable complex processing augmentations (to Types 2-4) as described in Section~\ref{sec:ACiS}. 

Of special mention is P4, the ``domain-specific programming language for network devices, specifying how data plane devices (e.g, switches, routers, NICs, filters, etc.) process packets'' \cite{P422}. P4 has aided in increasing switch flexibility \cite{Bosshart14} including accelerating applications that are beyond the basic switching function. Examples are consensus algorithms \cite{Dang16}, database transaction processing \cite{Jasny22}, caching \cite{Liu17}, and key-value store \cite{Jin17}. While there has also been some exploration of application-level processing \cite{Sankaran21} in P4 switches, including support of collectives in ML training \cite{Sapio21}, these capabilities are currently limited \cite{Swamy22} by the supported set of operations (e.g., no multiply), data types (e.g., no sparse data types), and memory footprint. Perhaps most significantly, packets can only access each memory location once within a traversal \cite{DeSensi21}; while it is possible to recirculate packets, this reduces throughput. To address these limitations, ACiS is designed to handle advanced calculations, such as fused multiply-accumulate (MAC) and sparse accumulation, off-chip memory access with cacheable buffers, data reuse, and control. 

In-switch computing is an active area of research with Types 1-3 (defined in Section~\ref{sec:ACiS}) being partially addressed. The work in \cite{YLi19} provides an in-switch computing paradigm, implemented on NetFPGA, to accelerate aggregation of gradients used in the training phase of reinforcement learning (Type 1).
The work in \cite{DeSensi21} designs a flexible programmable switch on top of PsPIN building blocks to accelerate Allreduce with custom operators and data types; that is, sparse data (Type 2 on a single collective type).
The work \cite{Liu20} presents an RDMA-compatible in-network reduction architecture to accelerate distributed DNN training in which the FPGAs are connected to the switch and the switch is configured to route the packets that need to be aggregated to the FPGA (Type 2).
The work in \cite{Swamy22} adds custom hardware based on a MapReduce pattern/abstraction (built upon a CGRA) to P4 devices to enable per-packet inference of machine learning (Type 3 - but with no fusion of collectives). To summarize, we are not aware of previous work that fully supports user-defined or complex collectives (Types 2 and 3) or in any way addresses look-aside capability (Type 4).

\section{Hardware Design}
\label{sec:hardware}

An important switch design consideration is modularity \cite{Michel21}, which ACiS follows by introducing \textit{composable plugins} to successively add capabilities. These plugins can be (nearly) seamlessly added to existing switches and are network friendly. Capabilities for Types 2-4 include flexible aggregation and communication management, schedulability, and stateful processing. 

Given that ACiS types are built in successive layers, the architecture is composed of successive plugins. Figure~\ref{fig:model-Type1to4} shows PISA, a state-of-the-art protocol-independent switch model, enhanced with an ACiS accelerator for Types 1-4. Only one pipe is shown in this figure; a switch is comprised of replicated pipes and each pipe can include a number of physical ports \cite{Verdi22}. We note that packet processing in this model is done at the header level. However, MPI-specific fields (e.g., source/destination ranks, tag) are embedded in the payload. Thus, we modify the architecture by introducing a separate pipeline for payload-level packet processing (bottom part of Figure~\ref{fig:model-Type1to4}). We summarize the key ideas for each plugin.

\begin{figure*}[htbp]
\centering
\includegraphics[width=0.8\linewidth]{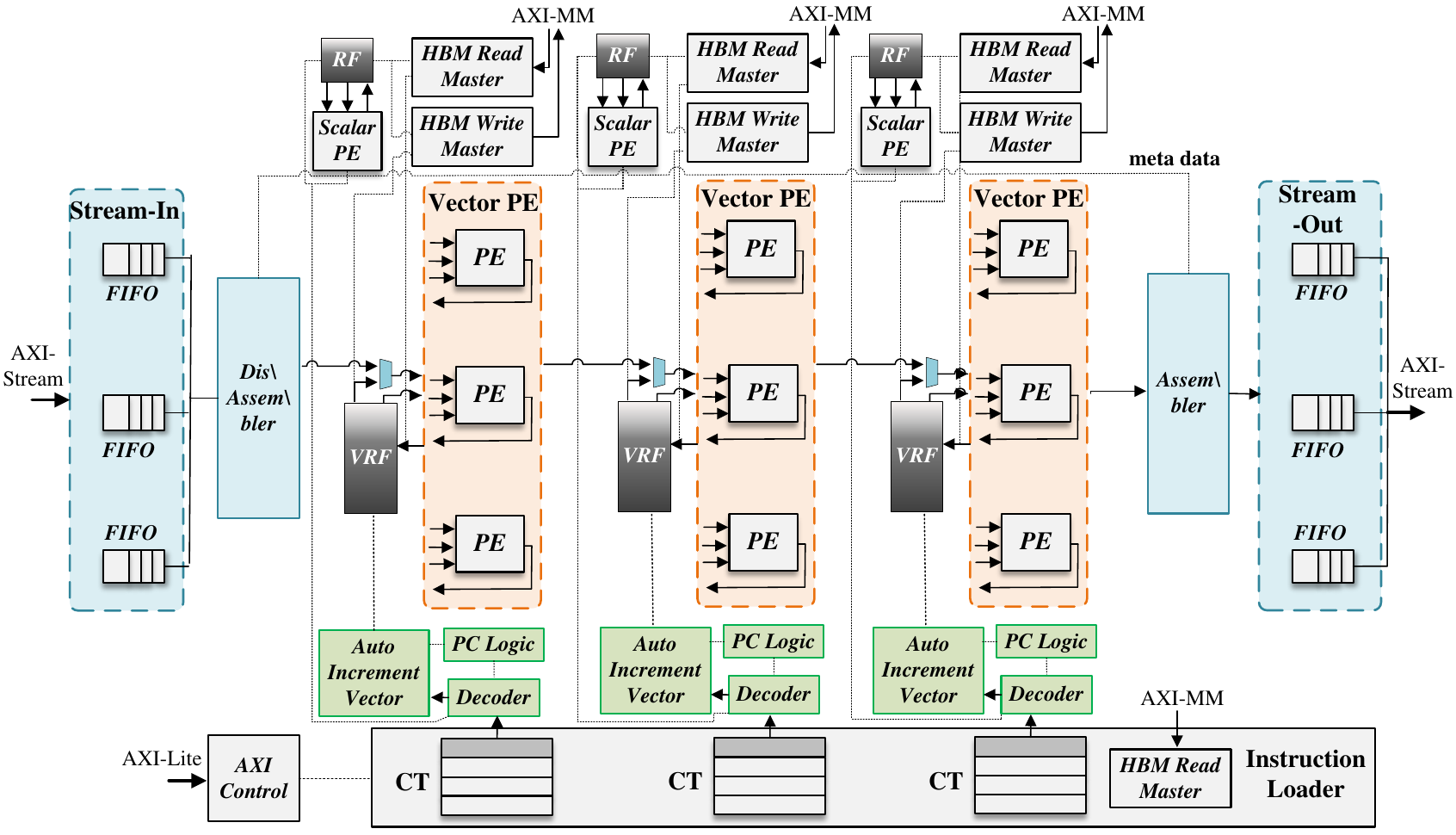}
\caption{\small The proposed CGRA architecture used in Figure \ref{fig:model-Type1to4} with three SIMD processing units (SPUs) in a deep pipeline. It is packaged with AXI interface to facilitate the integration with switch pipelines.}
\label{fig:cgra}
\vspace{-0.1truein}
\end{figure*}

\vspace*{0.04truein}
\noindent
{\bf Type 2} supports a collective processing unit including flexible management of MPI communicator context and programmable aggregation units with the support for different operations and data types. Two plugins are (1) a collective control unit, and (2) a programmable aggregation unit. We implement the collective control plugin with lookup tables. Current protocol-independent switches, however, do not offer flexible wide-access to these lookup tables; that restriction is addressed with this plugin. To route the output of the aggregation unit to any pipe (1-to-N converter) there is (3) a multicast engine at the end of the pipeline (inspired from the packet replication engine in P4 switches \cite{Shukla19}). Finally, in order to be compliant with any MPI implementation, we also need two other plugins: (4) a parser and (5) a deparser, but in the payload pipeline (see Figure \ref{fig:model-Type1to4}). These are again inspired from P4-based switches and make it feasible to support different MPI flavors. 

\vspace*{0.04truein}
\noindent
{\bf Type 3,} which supports state and loops, uses a plug-in that has an instruction-based reconfigurable compute unit and a recirculate interface (if the switch architecture does not already have one \cite{Ibanez19a}). The key plugin is the interface for off-chip memory, which facilitates stateful processing of HPC applications. To minimize the effort for the design of this type we note that it is equivalent to supporting load/store instructions in the CGRA (see Type 4 and Figure \ref{fig:cgra}). However, a good memory model is needed to fully exploit the capabilities of the hardware without hampering the overall application throughput. We take advantage of multi-banks provided by recent off-chip memory technologies and specify separate data and instruction memory for each SIMD Processing Unit (SPU).

\vspace*{0.04truein}
\noindent
{\bf Type 4,} Since the map operation (computation sandwiched between collectives) can be any user-provided function, the plugin must be programmable. We posit that a coarse grained reconfigurable array (CGRA) is a suitable candidate for the first plugin since both software-like programmability and near-ASIC performance is achieved. Using an instruction-capable CGRA makes the plugin user-friendly and brings the control plane closer to the switch fabric. We divide the CGRA's processing elements into SPUs and then stack these SPUs in a deep pipeline with a simple yet high speed interconnect. We observe that many HPC applications operate on vectors of data and these vectors are typically large. Consequently, the SPUs have wide vector instruction support. For the second plugin, a recirculate interface is used to process a chain of collectives.

\section{ACiS Experimental Evaluation}
\label{sec:results}

In this section, we present the experimental setup and the performance/scalability study of ACiS for collectives, HPC/AI kernels, and applications for different types.

\subsection{Experimental Setup}
\label{sec:setup}
For proof-of-concept, we have implemented and tested ACiS on an FPGA-based system in both direct and indirect network settings using the Xilinx \textit{Vitis} unified software platform. We compare ACiS with a high-end CPU cluster.

\textbf{Indirect Network:} The testbed is a two-node system on CloudLab \cite{Handagala21} with a Xilinx Alveo U280 FPGA attached to a Dell Z9100-ON switch (total of three nodes including host). A 100 Gbps switch interconnects all three nodes. Each process in the leaf nodes, two in this case, in addition to the FPGA itself, is assigned an IP address and a port number. This information is stored in the networking kernel of the FPGA to forward the messages to the correct destination according to the collective type and algorithm. Messages are sent from the leaf nodes to the FPGA through the switch, processed in the FPGA user kernel, and sent back to the corresponding leaf node(s). We also provide a runtime that automates and manages the execution of processes in basic/fused collectives. This includes connecting to leaf nodes from the master process (through SSH), creating processes there, assigning new port numbers for each process, and waiting for the completion. We use the Xilinx Vitis 2021.2 unified software platform to program the FPGA. Our accelerator is coupled with a modified version of \cite{xupudp} to send/receive packets from two leaf nodes. The operating frequency is 250 MHz. 

\textbf{Direct Network:} In the direct network testbed, two Alveo U280 boards are directly connected using QSFP28 network interfaces (capable of 100 \textit{Gb/s}). Each board is connected to an Intel Xeon E5-2620V2 server. 
To simulate a larger number of nodes, we conduct an experiment to obtain parameters used for the emulation of a larger-scale proxy system.
In this experiment, a sender process sends 1408 bytes worth of data to a receiver process using TCP/IP network logic \cite{He21} handled by FPGAs. 
We used the ExaMPI implementation \cite{Skjellum2020} as the middleware for this experiment. The parameters used in the emulation (derived from our system setup) are shown in Table \ref{tab:logcn}. MPI overhead is the average overhead of MPI\_send and MPI\_Recv in ExaMPI. 

For the rest of this section, we evaluate the performance of ACiS on the direct network based on the emulator. The emulation has the same requirements as \cite{Li2019}. That is, the emulation should possess: (1) the same volume of traffic in the network links, (2) an identical number of network hops, and (3) an accurate overhead of the accelerator. For the ACiS accelerator overhead, we use cycle-accurate RTL simulation through testbenches using the Xilinx Vivado Tool. We emulate an FPGA cluster with up to 128 nodes in a 3D-torus. 

\textbf{CPU Implementation:} For the CPU reference, we use the TACC Stampede2 \cite{Stanzione17} Skylake (SKX) compute cluster with 48-cores per node (2 sockets) 2.1~$GHz$ Intel Xeon Platinum 8160 CPUs, and a 100 Gb/s Intel Omni-Path (OPA) network. We used Intel MPI 18.0.2 as an Intel-compatible MPI is recommended for this cluster; we found it usually gives better performance than other MPI implementations.

\begin{table}[]
\caption{Parameters used in the emulation of the direct network. \\ * Denotes the Aurora IP latency.}
\vspace*{-0.05truein}
\label{tab:logcn}
\centering
\begin{tabular}{|c|c|}
\hline
MPI Overhead              & 14.8 usec \\ \hline
Maximum Network Bandwidth (BW) & 95.9 Gbps \\ \hline
PCIe Latency                   & 0.9 usec  \\ \hline
FPGA-to-FPGA Latency*           & 0.44 usec \\ \hline
Minimum Port-to-Port Latency   & 52 nsec   \\ \hline
\end{tabular}
\vspace*{-0.1truein}
\end{table}

\subsection{Performance and Scalability}

\begin{figure*}[htbp]
\centering
\includegraphics[width=1\linewidth]{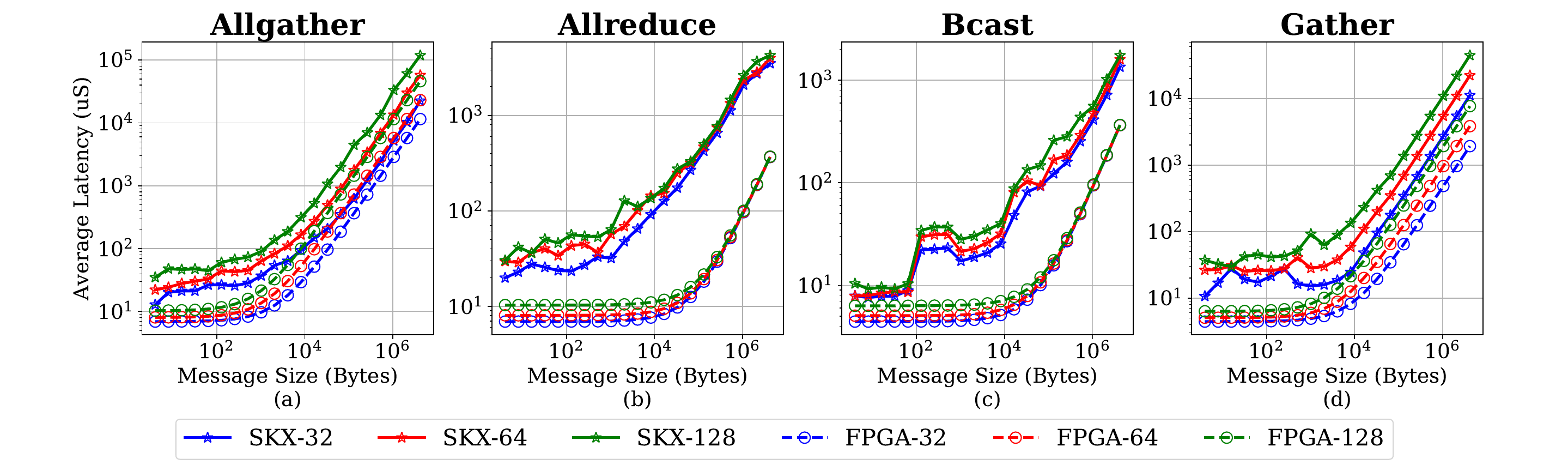}
\vspace*{-0.2truein}
\caption{ACiS vs MPI CPU cluster (SKX) execution times for 32, 64, and 128 nodes: (a) osu\_allgather, (b) osu\_allreduce, (c) osu\_bcast, and (d) osu\_gather.}
\label{fig:eval_merged}
\end{figure*}

\textbf{Type 2:}
Fig. \ref{fig:eval_merged} shows the simulation results of ACiS and baseline MPI collectives for small (4 Bytes to 4 KB) and medium-to-large message sizes (4 KB to 4 MB) for 32, 64, and 128 nodes (direct network) using the OSU benchmarks (v5.6.2) \cite{mvapich}. The reported average latency is the average time it takes for the processes to finish the operation. Processor-FPGA communication latency is included in the time. To isolate the impact of the design under study, e.g., from contention at the PCIe interface, we focused simulations with one process per node. As the results suggest, ACiS demonstrates greater performance compared to the CPU cluster baseline. 

\begin{figure*}[htbp]
    \centering
    \includegraphics[width=\linewidth]{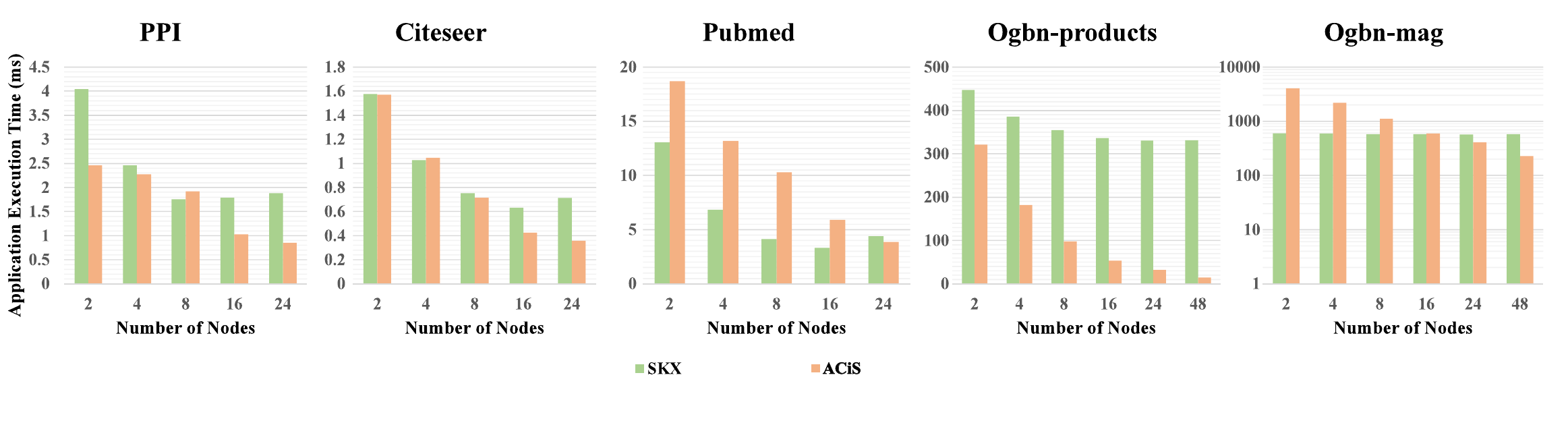}
    \vspace*{-0.5truein}
    \caption{Application performance and scalability comparison of GCN on a baseline CPU cluster (SKX) vs. ACiS.}
    \label{fig:app}
    \vspace*{-0.1truein}
\end{figure*}

\textbf{Type 3:}
We evaluate Type 3 for a Graph Convolution Network (GCN) application on the direct network using four datasets: PPI, Citeseer, Pubmed, ogbn-mag, and ogbn-products. Figure \ref{fig:app} shows the performance and scalability of GCN with and without ACiS acceleration across all datasets. It demonstrates superior scalability of ACiS. On average, ACiS improves application performance compared to a baseline SKX cluster by a factor of 2.2$\times$, 2$\times$, 1.1$\times$, 1.4$\times$, and 10.1$\times$ for PPI, Citeseer, Pubmed, ogbn-mag, and ogbn-products on 24 nodes with an average of 3.4$\times$ across all datasets. 

\begin{figure}[htbp]
\vspace*{-0.1truein}
\centering
\includegraphics[width=1.0\linewidth]{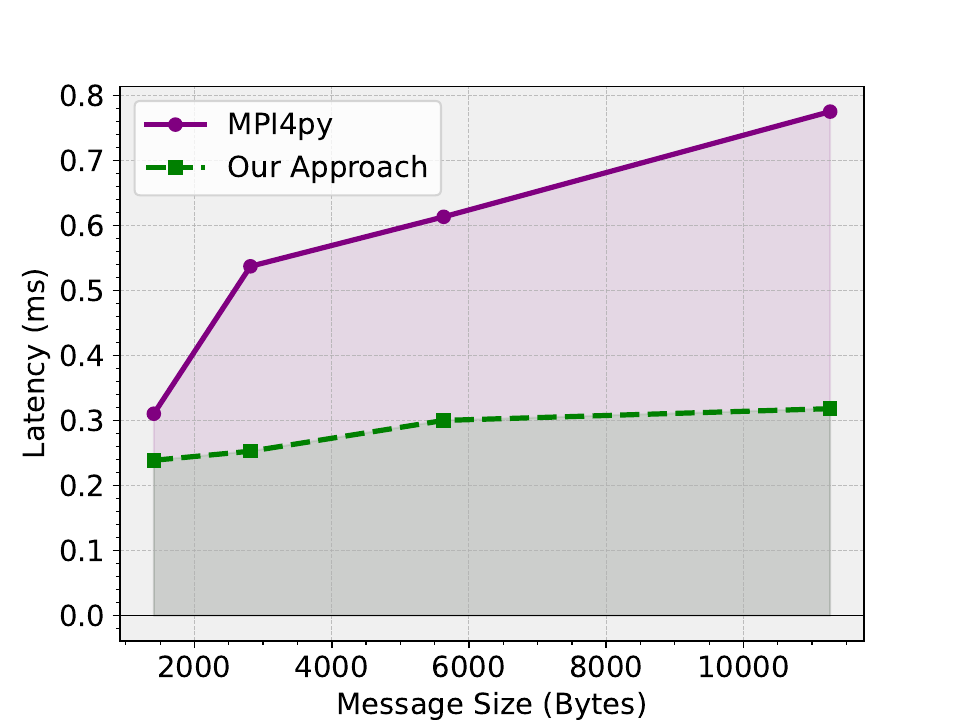}
\vspace*{-0.1truein}
\caption{Latency comparison of Allgather\_op\_Allgatherv in MPI4py and ACiS. \textit{Op} is prefix sum. The X-axis shows the message size in bytes used for Allgathers and the Y-axis shows the latency in milliseconds.}
\label{fig:indirect}
\vspace*{-0.1truein}
\end{figure}

\textbf{Type 4:}
\textbf{Indirect Network:}
To evaluate the efficacy in indirect network settings we use CloudLab  \cite{Handagala21} with three nodes (two leaf nodes and one node to host the FPGA) capable of 100 Gbps.
Since the runtime and MPI support are based on Python, we compare this approach with a Python-based MPI, MPI4py \cite{Dalcin21}. We demonstrate its performance compared to traditional MPI for an instance of fused collectives used in FEM applications. Figure \ref{fig:indirect} shows the latency comparison of Allgather\_op\_allgather in both MPI4py and ACiS for different message sizes. \textit{Op} here is a prefix sum. The results are from taking the average of five runs. It clearly shows that ACiS provided superior performance, especially for larger message sizes with, on average, a 1.98$\times$ improvement. The performance benefit comes from the fact that intermediate communications are bypassed and computations--sandwiched between collectives--are directly processed in the accelerator.

\begin{figure}[htbp]
\centering
\includegraphics[width=1.0\linewidth]{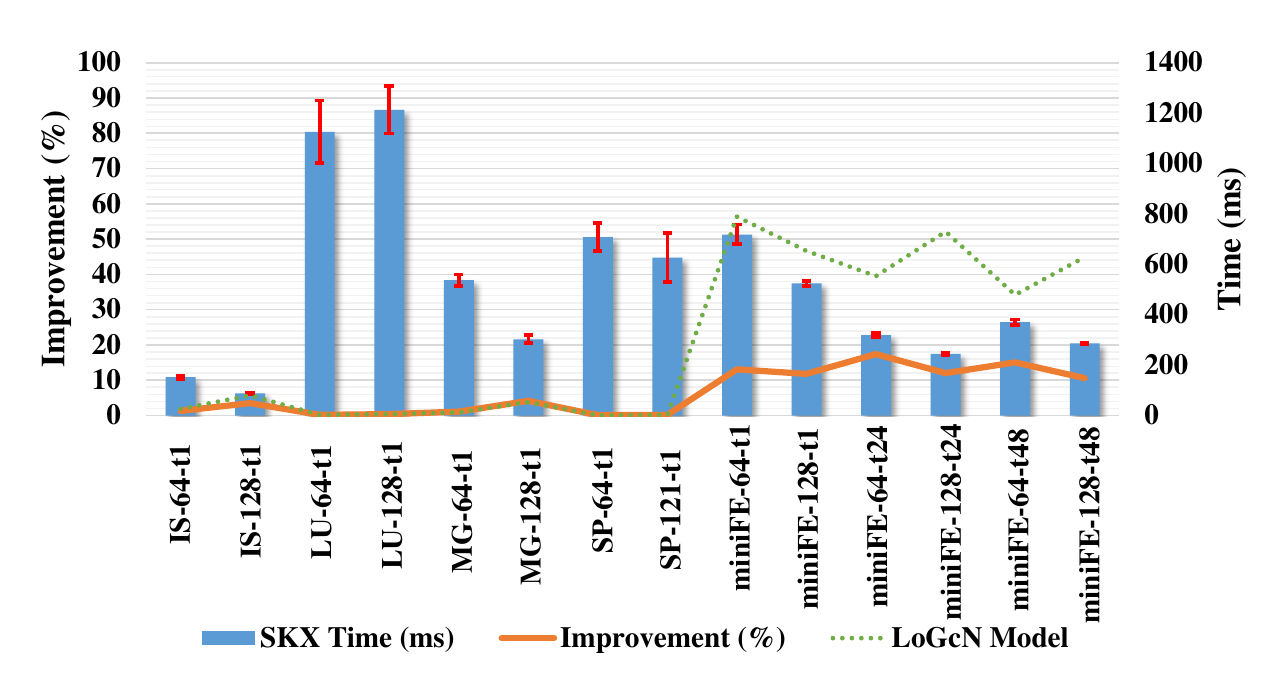}
\vspace*{-0.2truein}
\caption{Performance improvement of ACiS over original MPI implementation on SKX cluster (64 and 128 nodes) for the \textsc{NAS} parallel benchmarks and \textsc{miniFE}. SKX time and error bars represent the time it takes on SKX cluster and \textit{std} for five runs, respectively.}
\label{fig:mini_app}
\vspace{-0.1truein}
\end{figure}

\textbf{Direct Network:}
We evaluate ACiS against the baseline CPU cluster for NAS parallel benchmark (NPB) \cite{NAS} and miniFE \cite{minife}. NPB includes the following: IS (Integer Sort), LU (Lower-Upper Gauss-Seidel solver), MG (Multi-Grid on a sequence of meshes), and SP (Scalar Penta-diagonal solver). And miniFE is an unstructured implicit finite element code. Figure~\ref{fig:mini_app} shows the performance benefit of ACiS over the original MPI implementation on the SKX cluster (64 and 128 nodes) for the \textsc{NPB} and \textsc{miniFE} proxy applications. 
SKX time and error bars represent the time it takes on SKX cluster and \textit{std} for five runs. \textsc{Benchmark-X-tY} represents the benchmark with \textit{X} nodes and \textit{Y} OpenMP threads.
Among \textsc{NPB} applications, the performance benefits for \textsc{MG} and \textsc{IS} are higher than for the others. For \textsc{IS}, one reason is that the message size of collectives is relatively high, and ACiS can take advantage of communication-computation overlap and in-network data reduction. 
For \textsc{miniFE}, the performance improvement percentage is typically higher than that of \textsc{NPB}. 

\section{Software Support}
\label{sec:software}

Architectural enhancements should ideally be transparent to all other aspects of the system. This inevitably requires software support. Two aspects of this support are described, followed by a database that allows programmers to automatically find where ACiS collectives could collectives.

\subsection{MPI Transparency}

To support MPI transparently, and thus a large fraction of HPC applications, a new transport layer needs to be created using device-specific APIs to communicate with the switch accelerator device. Each supported MPI routine can then be modified to use the new transport (see, e.g., \cite{Haghi22}. We have already developed an FPGA transport to communicate between MPI processes through a secondary FPGA-directed network for an ExaMPI implementation \cite{Skjellum20}. ExaMPI is a light-weight MPI implementation, being developed by one of the PIs, which focuses on key blocks of functionality and new MPI concepts. 

\subsection{Configuring ACiS in the Switch}

To facilitate programming and interaction with a switch accelerator directly from an MPI application support for a source-to-source (S2S) translator should be added. The S2S translator includes (1) a parser, (2) a compiler, (3) an assembler, and (4) a wrapper. The input is the source code for an MPI application and the output is new MPI code enhanced with fused collective routines. New routines are added for each fused collective; each is modified with a loader (in addition to Types 1-2 software support) to load the CGRA binary (discussed below) into the switch accelerator. 

We now briefly describe its features \cite{Haghi23}. It is based on LLVM \cite{Lattner2004} with the back-end target instruction set being RISC-V (used by the CGRAs).
LLVM emits an intermediate representation (IR). 
In the back-end, target-dependent parameters (e.g., CGRA dimension) are used to apply a number of optimizations. The steps in the back-end include parsing the IR, generating the DFG, code generation, scheduling, and register allocation. Next, an assembler takes the generated instructions and outputs CGRA binary. Finally, the wrapper replaces parts of the application containing the fused collectives with the corresponding fused collective routine. The generated CGRA binary is carried as an argument to the fused collective routine.

\subsection{Database}

In \cite{Haghi23a} we characterize a large number of MPI applications to determine {\it overall} applicability of ACiS-supported operations, in both breadth and type, and so provide insight for hardware designers and MPI developers about future offload possibilities. Besides increasing the scope of prior surveys to include finding (potential) new MPI constructs, we also tap into new methods to extend the survey process. Prior surveys on MPI usage considered lists of applications constructed based on application developers' knowledge. The we take, however, is based on an automated \textit{mining} of a large collection of code sources. More specifically, the mining is accomplished by GitHub REST APIs. We use a database management system to store the results and to answer queries. Another advantage is that this approach provides support for a more complex analysis of MPI usage, which is accomplished by user queries.

\section{Conclusion}
\label{sec:conclusion}

In this work, we propose a general-purpose, transparent, framework for in-switch computing in reconfigurable devices to provide application-level acceleration. We concentrate on communication collectives and their combinations with each other and with mapping functions. We find that the same basic hardware substrate can be used for multiple extensions to basic collectives: user definition, look-aside (context), and fusion (including some level of control). We describe some of the software support that enables transparency.

\section*{Acknowledgements}

This work was supported, in part, by the National Science Foundation through awards CCF-1919130, CCF-2151021, and CCF-2326494; and by AMD and Intel both through donated FPGAs, tools, and IP.

\def\authornoop#1{}


\end{document}